\documentstyle[epsf,onecolumn]{mn}

\title[HR 1217 - a new frequency]{Discovery of the ``missing'' mode
 in HR 1217 by the Whole Earth Telescope}

\author[D. Kurtz et. al. (the WET collaboration)]{
D. Kurtz$^{1, 2}$,
S.D. Kawaler$^3$, R.L. Riddle$^3$, M.D. Reed$^{3, 4}$, M.S. Cunha$^{5}$,\cr
N. Silvestri$^{6}$, 
M. Wood$^{6}$, T.K. Watson$^{7}$, 
N. Dolez$^{2}$, P. Moskalik$^{8}$, 
S. Zola$^{9}$, \cr
E. Pallier$^{2}$, 
J.A. Guzik$^{10}$, T.S. Metcalfe$^{11}$, A. Mukadam$^{11}$, 
R.E. Nather$^{11}$,  \cr
D.E. Winget$^{11}$,
D.J. Sullivan$^{12}$,
T. Sullivan$^{12}$, K. Sekiguchi$^{13}$,
X. Jiang$^{14}$,\cr
R. Shobbrook$^{15}$,
B.N. Ashoka$^{16}$,
S. Seetha$^{16}$,
S. Joshi$^{17}$,
D. O'Donoghue$^{18}$, \cr
G. Handler$^{18}$, 
M. Mueller$^{18}$,
J.M. Gonzalez Perez$^{19}$, 
J.-E. Solheim$^{19}$, \cr
F. Johannessen$^{19}$, 
A. Ulla$^{20}$,
S.O. Kepler$^{21}$, A. Kanaan$^{22}$, A. da Costa$^{21}$, \cr
L. Fraga$^{22}$, O. Giovannini$^{23}$,
and J.M.~Matthews$^{24}$
\\
$^{1}$Centre for Astrophysics, University of Central Lancashire,
 Preston PR1 2HE, UK,\\
 Department of Astronomy, University of Cape Town, Rondebosch 7701, 
South Africa\\
$^{2}$Observatoire Midi-Pyr\'{e}n\'{e}es,
 CNRS/UMR5572, 14 av. E. Belin, 31400 Toulouse, France\\
$^{3}$Department of Physics and Astronomy, Iowa State University, Ames, 
IA  USA\\
$^{4}$Visiting Astronomer, CTIO, National Optical
Astronomy Observatories\\
$^{5}$Centro de Astrof\'\i sica da Universidade do Porto, Rua das 
Estrelas, 4150-762 Porto, Portugal\\
\& Instituto Superior da Maia, Lugar de vilarinho, 4470
Castelo da Maia, Portugal\\
$^{6}$Department of Physics \& Space Sciences and SARA Observatory, Florida \\ 
 Institute of Technology, Melbourne, FL, USA\\
$^{7}$Southwestern University, 1001 E. University Avenue, Georgetown,
 TX, USA\\
$^{8}$Nicolas Copernicus Astronomical Centre, ul. Bartycka 18, 00-716
 Warsawa, Poland\\
$^{9}$Krakow Pedagogical University, ul.
 Podchor\c{a}\.{z}ych 2, Krakow, Poland\\
$^{10}$X-2, MS B220, Los Alamos National Laboratory, Los Alamos, NM  87545  USA\\
$^{11}$McDonald Observatory, and Department of Astronomy, University 
of Texas,
 Austin, TX 78712, USA\\
$^{12}$School of Chemical and Physical Sciences, Victoria University 
of Wellington, PO Box 600, Wellington, New Zealand\\
$^{13}$Subaru Observatory, NAOJ, Hilo, HI USA\\
$^{14}$National Astronomical Observatories and Joint Laboratory of 
Optical Astronomy,\\ Chinese Academy of Sciences, Beijing, 100012, China\\
$^{15}$Research School of Astronmomy \& Astrophysics, 
Australian National University, Cotter Road, Weston, ACT 2611,  
Australia\\
$^{16}$Indian Space Research Organization, Vimanapura PO, 
 Bangalore 560 017, India\\
$^{17}$State Observatory, Manora Peak, Naini Tal 263 129, India\\
$^{18}$South African Astronomical Observatory, PO Box 9, Observatory 
7935, South Africa\\
$^{19}$Department of Physics, University of Tromso, N-9037 Tromso, Norway\\
$^{20}$Depto. de Fisica Aplicada, Universidade de Vigo,
 36200 Vigo, Spain\\
$^{21}$Instituto de Fisica, UFRGS, Campus do Vale, C.P. 15051,
Porto Alegre, RS, Brazil\\
$^{22}$Universidade Federal de Santa Catarina, Florian\'opolis, SC - 
Brazil\\
$^{23}$Departamento de F\'{\i}sica e Qu\'{\i}mica, Universidade de Caxias do
Sul, 95001-970 Caxias do Sul, RS - Brazil\\
$^{24}$Dept. of Physics and Astronomy, University of British Columbia, 
Vancouver, Canada
}

\begin{document}
 
\maketitle
\twocolumn

\begin{abstract}
HR\,1217 is a prototypical rapidly oscillating Ap star that has
presented a test to the theory of nonradial stellar pulsation.  Prior
observations showed a clear pattern of five modes with alternating
frequency spacings of 33.3\,$\mu$Hz and 34.6\,$\mu$Hz, with a sixth
mode at a problematic spacing of 50.0\,$\mu$Hz (which equals $1.5
\times 33.3 \mu$Hz) to the high-frequency side.  Asymptotic pulsation
theory allowed for a frequency spacing of 34\,$\mu$Hz, but {\sc
hipparcos} observations rule out such a spacing.  Theoretical
calculations of magnetoacoustic modes in Ap stars by Cunha (2001)
predicted that there should be a previously undetected mode
34\,$\mu$Hz higher than the main group, with a smaller spacing between
it and the highest one.  In this Letter, we present preliminary
results from a multi-site photometric campaign on the rapidly
oscillating Ap star HR\,1217 using the ``Whole Earth Telescope''. 
While a complete analysis of the data will appear in a later paper,
one outstanding result from this run is the discovery of a newly
detected frequency in the pulsation spectrum of this star, at the
frequency predicted by Cunha (2001).
\end{abstract}

\begin{keywords}
Stars: oscillations -- stars: variables -- stars: individual 
(HR\,1217) --
stars: magnetic.
\end{keywords}

\section{Introduction}

After decades of trying, the search for solar-type oscillations in
stars finally appears to have been successful (see, e.g., Bouchy \&
Carrier 2001 and Carrier et al.  2001).  Although this led Gough
(2001) to announce the ``birth of asteroseismology'', for the past two
decades asteroseismology has successfully investigated the interiors
of many types of stars other than the solar-type stars.  Remarkable
success stories of observational and theoretical investigations of
white dwarf stars and rapidly oscillating Ap stars have amply
demonstrated the power of asteroseismology as a tool to advance our
knowledge of the physics of stellar interiors and the details of
stellar evolution (see, e.g., Kurtz et al 1989; Winget et al.  1991;
Kawaler \& Bradley 1994; Matthews et al. 
1999).

With these successes, some mysteries have remained.  In this Letter,
we address apparently contradictory interpretations of the pulsation
spectrum of the rapidly oscillating Ap star HR\,1217.  This star,
discovered to be a pulsator by Kurtz (1982), was investigated with an
extensive global campaign in 1986 (Kurtz et al.  1989).  A key
result from that data set was a list of six principal pulsation
frequencies, reproduced in Table 1.  As expected from the asymptotic
theory of nonradial pulsations, five of the modes are nearly equally
spaced in frequency.

The asymptotic frequency spacing, $\nu_o$, is a measure of the sound
crossing time of the star, which in turn is determined by the star's
mean density and radius.  With a typical mass of Ap stars of about
2M$_{\odot}$, $\nu_o$ reflects the radius of the star, with $\nu_o$
scaling as $R^{-3/2}$.  In the asymptotic limit, the number of nodes
in the radial direction, $n$, is larger than the spherical degree
$\ell$.  Assuming adiabatic pulsations in spherically symmetric stars
the pulsation frequencies are, to first order,
\[
\nu_{n,\ell} = \nu_o (n + \ell/2 + \epsilon),
\]
where $\epsilon$ is a (small) constant (Tassoul 1980, 1990). Without 
precise identification of the degree ($\ell$) of the pulsation modes, 
asymptotic theory allows the frequency spacing to be uncertain by a 
factor of two, depending on whether modes of alternating even and odd 
$\ell$ are present (producing modes separated by $\nu_{o}/2$ in 
frequency), or only modes of consecutive $n$.

The results of the 1986 campaign were inconclusive as to whether
$\nu_o$ was 68\,$\mu$Hz or 34\,$\mu$Hz.  The prinicipal frequencies
seen in the data are given in Table 1; they correspond to those found
by Kurtz et al.  (1989) for the 15 day stretch of best coverage (for
comparison with the new data presented in Table 3).  The highest
frequency of HR\,1217 in those data was 50\,$\mu$Hz higher than the
fifth mode, suggesting that $\nu_o$ was 34\,$\mu$Hz.  But the fine
structure of the spacings was suggestive of alternating $\ell$ values. 
Fortunately, the two possible values could be assessed if the
luminosity of the star were precisely known.  If $\nu_o$ were
34\,$\mu$Hz, then the radius of HR\,1217 would be large enough that it
would be far removed from the main sequence (i.e. more evolved) and
therefore more luminous (Heller \& Kawaler 1988).  Matthews et al. 
(1999) used the {\sc hipparcos} parallax measurement to place HR\,1217
unambiguously close to the Main Sequence -- meaning that $\nu_o$ is
indeed 68\,$\mu$Hz.  This deepened the ``mystery of the sixth
frequency'', now $\frac{3}{4}\nu _0 $ higher.  No clear theoretical
construct could explain it.

The asymptotic frequency spacing given in the equation above is valid
only for linear adiabatic pulsations in spherically symmetric stars. 
However, the magnetic field, the chemical inhomogeneities, and
rotation all contribute to break the spherical symmetry in roAp
stars.  Therefore, it is important to know the effects that these
deviations from spherical symmetry have on the theoretical amplitude
spectra of roAp stars, before comparing the latter with the observed
amplitude spectra.  The effects of the chemical inhomogeneities have
been discussed recently by Balmforth et al.  (2001), but those will
not concern us further here.  The effects of the magnetic field on the
oscillations of roAp stars (Dziembowski \& Goode 1996; Bigot et al. 
2000; Cunha \& Gough 2000), as well as the conjoined effect of
rotation and magnetic field (Bigot 2002), have been determined by
means of a singular perturbation approach.  While generally the
magnetic field effect on the oscillations is expected to be small,
Cunha \& Gough (2000) found that, at the frequencies of maximal
magnetoacoustic coupling, the latter is expected to become
significantly large, resulting in an abrupt drop of the separation 
between mode frequencies.

The observational consequence of the results of Cunha \& Gough (2000)
suggests that we should see equally spaced modes in roAp stars, with
an occasional mode much closer to its lower frequency counterparts. 
More recently, Cunha (2001) suggested that the explanation of the
strange separation between the last two modes observed in HR\,1217
could rest on the occasional abrupt decrease of the large separations
predicted by Cunha \& Gough (2000).  For this prediction to hold, she
argued that the observations of Kurtz et al.  (1989) must have missed
detecting a mode at a frequency 34$\mu$Hz higher than that of the
fifth mode they observed.  She predicted that new, more precise
measurements would find this ``missing mode'' if the Alfv\'enic
losses were not large enough to stabilise it.  Detailed
re--examination of the data from 1986 shows no peak at the key
position approximately 33 $\mu$Hz above f5 at the 0.1~mma level.

\begin{table}
\caption{Principal frequencies in HR\,1217 (data from Kurtz et al. 
1989).}
\begin{tabular}{cccc}
Number & frequency & frequency spacing & amplitude \\
       & [$\mu$Hz] &  [$\mu$Hz]        & [mmag] \\

f1 & 2619.51$\pm 0.05$ &    -            & 0.28$\pm 0.03$\\
f2 & 2652.92$\pm 0.02$ & 33.41$\pm 0.05$ & 1.09$\pm 0.03$ \\
f3 & 2687.58$\pm 0.03$ & 34.66$\pm 0.04$ & 0.94$\pm 0.03$ \\
f4 & 2721.02$\pm 0.02$ & 33.44$\pm 0.04$ & 1.16$\pm 0.03$ \\
f5 & 2755.49$\pm 0.04$ & 34.47$\pm 0.04$ & 0.49$\pm 0.03$ \\
f6 & -- & -- &  $< 0.09$ \\
f7 & 2806.26$\pm 0.06$ & 50.77$\pm 0.07$ & 0.25$\pm 0.03$ \\

\end{tabular}
\end{table}

In 2000 November, we began an extensive, coordinated global photometry
campaign on HR\,1217 using the Whole Earth Telescope.  A complete
analysis of this extensive data set, which addresses many other
aspects of roAp stars, is in preparation.  In this Letter, we present
a preliminary analysis of data from that run that clearly shows a
previously unseen pulsation mode at a frequency about 36\,$\mu$Hz
above the fifth frequency, as predicted by Cunha (2001).  In the next
section, we describe the observational procedures and the data
coverage and reduction.  Section 3 presents the preliminary frequency
analysis, and the results are discussed in Section 4, along with a
brief discussion of the impact of this result on the theory of
pulsations of roAp stars.

\section{Observations}

The WET run on HR\,1217 began on 2000 November 6 at selected sites, 
and 
continued through early 2000 December. The bulk of the data, with the 
best global coverage, were obtained during 2000 November 14-30. A 
complete analysis of all of the available data is 
currently underway. For this Letter, we concentrate on the central 
portion of the WET run, with data from five sites. This subset of the 
full data set provides high signal--to--noise and a reasonable global 
coverage. It also extends over slightly more than one rotation cycle 
of HR\,1217. Since the pulsation amplitude is modulated with the 
rotation 
period, this data subset is just long enough to begin to resolve 
rotational sidelobes of the main peaks.

Table 2 lists the individual observing runs in this data set.  The
telescopes used range in aperture from 0.6\,m to 2.1\,m.  Data from
all sites were obtained using photoelectric photometers, with 10\,s
individual integrations.  At Beijing Astronomical Observatory,
McDonald Observatory, Mauna Kea Observatory, and Observatorio del
Teide, the observers used three-channel photometers that are
functionally similar to the equipment described in Kleinman et al. 
(1996).  The South African Astronomical Observatory observations were
made with a single-channel photometer, and the observations at CTIO
with a two--channel photometer.  At all sites, observations were made
through a Johnson $B$ filter, along with neutral density filters when
needed to keep the count rates below $10^{6}$\,s$^{-1}$.  Following
the procedures described in Kleinman et al.  (1996), the sky
background was continuously monitored with the three-channel
instruments.  At sites using two-channel and single channel
photometers, the sky was obtained several times during the night at
irregular intervals, and then interpolated during reduction.

\begin{table}
\caption{Observing log of selected high-speed photometry of HR\,1217 
from the Whole Earth Telescope Extended Coverage Campaign 20 (WET 
Xcov20)}
\begin{tabular}{lccclc}
	
Run Name & Date & Start & Run & Observatory & Tel \\
 & 2000 & (UT) & (hr)& & (m) \\
 & & & & & \cr
sa-od044 & Nov 14 & 21:03:00 & 5.04 & SAAO & 1.9 \\
mdr-142 & Nov 15 & 01:28:10 & 5.06 & CTIO & 1.5 \\
sa-od045 & Nov 15 & 19:20:00 & 7.05 & SAAO & 1.9 \\
teide01 & Nov 16 & 00:42:10 & 3.56 & Teide & 0.8 \\
mdr-143 & Nov 16 & 01:23:00 & 7.23 & CTIO & 1.5 \\
no1700q2 & Nov 17 & 07:28:00 & 3.38 & Mauna Kea & 0.6 \\
mdr-144 & Nov 17 & 20:26:06 & 7.54 & CTIO & 1.5 \\
teiden04 & Nov 17 & 22:09:10 & 6.05 & Teide & 0.8 \\
no1800q1 & Nov 18 & 07:22:30 & 4.25 & Mauna Kea & 0.6 \\
teiden05 & Nov 18 & 22:53:20 & 5.40 & Teide & 0.8 \\
sa-od047 & Nov 18 & 23:29:00 & 1.45 & SAAO & 1.9 \\
no1900q2 & Nov 19 & 10:14:20 & 3.85 & Mauna Kea & 0.6 \\
sa-od048 & Nov 19 & 18:55:00 & 7.15 & SAAO & 1.9 \\
teiden06 & Nov 19 & 22:05:30 & 6.06 & Teide & 0.8 \\
no2000q1 & Nov 20 & 07:37:00 & 6.07 & Mauna Kea & 0.6 \\
sa-od049 & Nov 20 & 18:51:00 & 7.30 & SAAO & 1.9 \\
sa-m0003 & Nov 21 & 19:26:50 & 6.67 & SAAO & 0.75 \\
sa-m0004 & Nov 22 & 18:28:20 & 7.65 & SAAO & 0.75 \\
no2300q1 & Nov 23 & 07:15:50 & 4.59 & Mauna Kea & 0.6 \\
teiden10 & Nov 23 & 22:05:40 & 5.47 & Teide & 0.8 \\
sa-m0006 & Nov 24 & 18:18:00 & 7.76 & SAAO & 0.75 \\
no2500q1 & Nov 25 & 07:03:00 & 6.67 & Mauna Kea & 0.6 \\
teiden12 & Nov 25 & 22:09:20 & 5.61 & Teide & 0.8 \\
joy-012 & Nov 26 & 03:55:50 & 4.10 & McDonald & 2.1 \\
no2600q2 & Nov 26 & 06:59:30 & 6.47 & Mauna Kea & 0.6 \\
sa-m0007 & Nov 26 & 18:28:40 & 7.42 & SAAO & 0.75 \\
no2700q1 & Nov 27 & 06:38:00 & 5.55 & Mauna Kea & 0.6 \\
jxj-0127 & Nov 27 & 13:44:10 & 4.75 & Beijing AO & 0.85 \\
sa-m0008 & Nov 27 & 18:27:50 & 7.57 & SAAO & 0.75 \\
teiden14 & Nov 27 & 22:28:20 & 3.72 & Teide & 0.8 \\
sa-h-046 & Nov 28 & 18:54:30 & 6.41 & SAAO & 1.9 \\
teiden15 & Nov 28 & 22:01:50 & 5.52 & Teide & 0.8 \\
no2900q1 & Nov 29 & 06:41:00 & 6.77 & Mauna Kea & 0.6 \\
sa-gh465 & Nov 29 & 20:30:30 & 5.14 & SAAO & 1.9 \\
teiden16 & Nov 29 & 21:18:50 & 2.59 & Teide & 0.8 \\
joy-028 & Nov 30 & 03:54:20 & 5.24 & McDonald & 2.1 \\
no3000q1 & Nov 30 & 06:40:50 & 6.77 & Mauna Kea & 0.6 \\
sa-gh466-9 & Nov 30 & 19:30:20 & 6.26 & SAAO & 1.9 \\
 & & & & & \\
 & & Total & 215.2 & & \\

\end{tabular}
\end{table}

As can be seen in Table 2, we obtained 215.2 hr of observations 
during the interval from 2000 November 14--30, resulting in 
a duty cycle of 53\%. Longitude coverage was adequate, though the 
longitudes around central Asia were not as well covered as the 
others.

\section{Frequency analysis}

The Fourier transform (FT) of the reduced data, in the frequency range 
where the pulsations are significant, is shown in in the top
panel of Figure 1.  The spectral window, shown in the middle panel of
the figure, shows the response of the FT to a single, noise-free
sinusoid sampled at the same times as the light curve of HR\,1217. 
The side peaks correspond to aliases of 1\,d$^{-1}$ and 2\,d$^{-1}$. 
They are approximately 40\% of the amplitude of the principal peak, 
and are caused by the (small) daily gaps present in the data causing 
incomplete global coverage.

\begin{figure}
\epsfysize=11cm \epsfbox{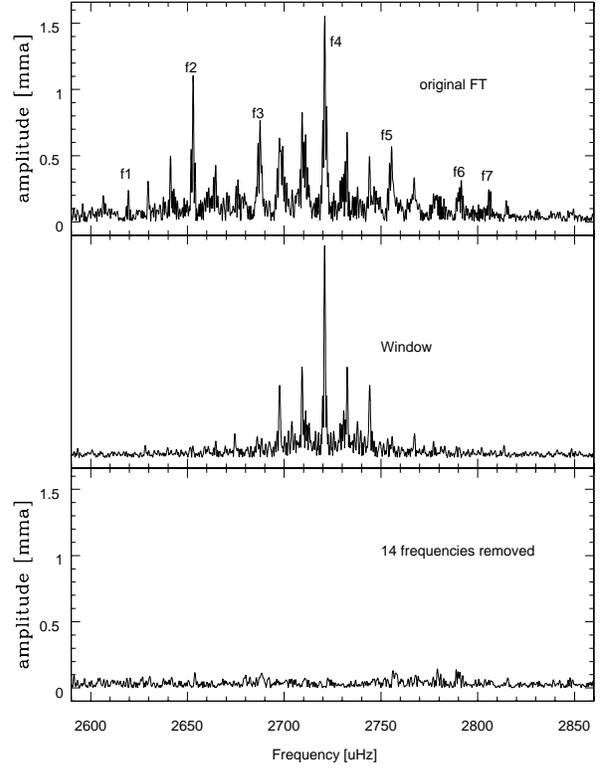}
\caption{The Fourier transform of the subset of
WET data used in our analysis. The top panel shows the FT of the
data, the middle panel shows the spectral window, and the bottom panel
shows the resulting FT after the data are prewhitened by 14 
frequencies. The unit mma means milli-modulation amplitude which is 
measured in parts per thousand in intensity units. For amplitudes as 
small as these here, it is very nearly equivalent to mmag.}
\end{figure}

The principal periodicities that we found in HR\,1217 are listed in
Table 3. Following initial identification of the main peaks in the
FT, we did a successive least-squares fit to the light curve including
all of the main peaks. We then included the rotational sidelobes in
the fit, sequentially. Throughout this process we prewhitened the
data by removing noise-free sinusoids at the fitted frequencies,
amplitudes, and phases. We stopped when none of the remaining peaks
was above the noise level. In all, we found 14 significant
periodicities in this data set. In addition to the 7 principal
frequencies, both rotational sidelobes of f3 and f4 were found. We
also found the low-frequency rotational sidelobe of f2, f5, and f7.
The frequencies listed in Table 3 are from the fit that included all 
14 frequencies.

The bottom panel of Fig.  1 shows the FT of the residual light curve
following the removal of 14 frequencies, on the same scale as the top
panel.  There are some residual peaks in this plot at interesting
frequencies.  Analysis of the full data set, including runs outside of
the subset that we used, shows that some of these are real.  They will
be described in further detail in the full analysis of the data which
is in preparation.

\begin{table}
\caption{Principal frequencies in HR\,1217 in 2000}
\begin{tabular}{cccc}
 
Number & frequency & frequency spacing & amplitude\\
 & [$\mu$Hz]& [$\mu$Hz] & [mma] \cr
& & & \cr
f1 & 2619.51 $\pm 0.03$ & - & 0.24 $\pm$0.02 \\
f2 & 2652.96 $\pm 0.01$ & 33.45 $\pm 0.04$ & 0.95 $\pm$0.02 \\
f3 & 2687.58 $\pm 0.02$ & 34.62 $\pm 0.02$ & 0.68 $\pm$0.03 \\
f4 & 2720.96 $\pm 0.02$ & 33.38 $\pm 0.03$ & 1.29 $\pm$0.02 \\
f5 & 2755.35 $\pm 0.03$ & 34.39 $\pm 0.04$ & 0.34 $\pm$0.02 \\
f6 & 2791.48 $\pm 0.03$ & 36.13 $\pm 0.04$ & 0.29 $\pm$0.02 \\
f7 & 2806.43 $\pm 0.14$ & 14.95 $\pm 0.14$ & 0.22 $\pm$0.07 \\

\end{tabular}
\end{table}

\section{Results}

\subsection{Comparison with the 1989 data}

Early in the run, it became clear that HR\,1217 was pulsating with the
same frequencies that were present in the 1986 data analysed by Kurtz
et al.  (1989).  Tables 1 and 3 show that the principal frequencies
from the 1986 study (f1 through f5 and f7) are highly consistent over
a time span of 15 yr.  Some of the amplitudes of these modes are
higher in 2000 and some lower by small amounts than they were in 1986,
but it is the frequencies (and presence or absence) of the modes that
are of interest here.

\begin{figure}
\epsfysize=11cm \epsfbox{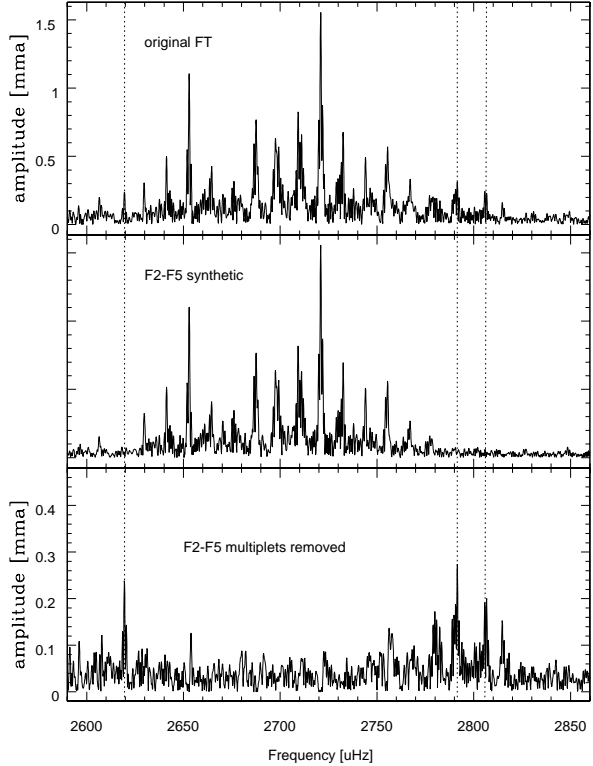}
\caption{The Fourier transform of the subset of
WET data used in our analysis. The top panel shows the FT of the
data. The middle panel is a simulation of the FT that includes 10 
frequencies - f2 through f5 along with their rotational sidelobes. 
The bottom panel shows the FT of the data prewhitened by those 10 
frequencies. Vertical dotted lines show the position of f1, f6, and 
f7.}
\end{figure}

The chief difference between the 2000 data and 1986 data is the
presence of a frequency at 2791 $\mu$Hz listed as f6 in Table 3. 
That mode was not detected in the data of Kurtz et al.  (1989 - Table
1) but was a clear signal in the WET Xcov20 2000 data.  To ensure that
this frequency is not an artefact of the data reduction algorithm, we
repeated the frequency analysis of our data fitting just the
large-amplitude peaks f2, f3, f4, and f5, and their rotational
sidelobes (if present).  We then removed those 10 frequencies.  The
results are illustrated in Fig.  2.  This figure shows the original
FT, and the FT of the data simulated by including f2-f5 and their
rotational sidelobes.  Clearly, there is excess signal at the positions
of f1 and f7, but also at 2791~$\mu$Hz as well.

Thus we conclude that the ``new'' frequency, f6, is real.  Table 3
shows that it lies at nearly $\nu_{o}/2$ above f5, as expected if it
is a normal $p-$mode and $\nu_{o}\approx 68\mu$Hz.  It is much closer
to f7 than $\nu_{o}/2$, as predicted by Cunha (2001).

\subsection{Implications for roAp stars}

Cunha (2001) speculated that the position of the f7 peak in the Kurtz 
et al. (1989) data is consistent with her model of the normal mode 
structure in Ap stars when magnetic fields are important to the 
pulsation dynamics. Since that peak was $\frac{3}{4}\nu _0 $ above f5 
(which is inexplicable in asymptotic theory), she suggested 
that there should be a peak at $\nu_{o}/2$ above f5. That is 
precisely what we see in the data from the WET run in November 2000 
with the discovery of f6.

As discussed earlier, other explanations for the frequency spacing
pattern from f5 to f7 are in direct conflict with the now
well-determined {\sc hipparcos} luminosity of HR\,1217.  We therefore
conclude that the frequency pattern in HR\,1217 suggests that the
pulsations we see in this star are consistent with normal $p-$mode
pulsations whose frequencies are, in some cases, strongly affected by
the magnetic field of the star.

Cunha (2001) suggested that large Alfv\'enic losses could help
explain the missing f6 in the 1986 data, as these losses are
maximal at the frequencies where the large separations experience the
abrupt decrease.  This energy loss could either stabilize the mode or
contribute to decrease its amplitude (although it is not clear how the
growth rates relate to the amplitude of the modes in roAp stars).

Since f6 is observed in the present data, the possibility that the
Alfv\'enic losses are large enough to stabilise this mode can be ruled
out, at least at the time of these observations.  Whether at the time
of the previous observations the efficacy of the magnetoacoustic
coupling (which depends, among other things, on the exact frequency of
the mode and on the characteristics of the magnetic field) was
different, is something to which we do not have an answer.  An attempt
to monitor the amplitude of f6, as well as that of the other modes, in
the future might, therefore, be worthwhile.  However, the magnetic
field does produce an important observable effect on the frequency of
f7.

With the detection of f6, we move closer to a detailed understanding
of the pulsation mechanism in roAp stars.  Most intriguingly, this
result for HR1217 suggests that, with appropriately detailed models,
we may soon be able to probe the magnetic field structure below the 
surfaces of these stars through their pulsation frequencies --
another application of asteroseismology to probing stellar interiors.

\section*{Acknowledgments}
We gratefully acknowledge support from the U.S. National Science Foundation
through grant AST-9876655 to Iowa State University, and funding by
UNESCO through the International Institute of Theoretical and Applied
Physics at Iowa State.  M.C is supported by FCT-Portugal through the
grant PD/18893/98 and the grant POCTI/1999/FIS/34549 approved by FCT
and POCTI, with funds from the European Community programme FEDER.
PM is supported by KBN (Poland) through grant 5-P03D-012-20.


\begin{thebibliography}{}

\bibitem[]{}Balmforth, N.J., Cunha, M.S., Dolez, N., Gough, D.O., 
Vauclair, S., 2001, MNRAS, 323, 362

\bibitem[]{}Bigot, L., 2002, in Radial and Nonradial Pulsations as 
probes of Stellar Physics, IAU colloquium 185, eds. C. Aerts, J. 
Christensen-Dalsgaard and T. Bedding, ASP Conf. Ser., in press

\bibitem[]{}Bigot, L., Provost, J., Berthomieu, G., Dziembowski, 
W.A., Goode, P.R., 2000, A\&A, 356, 218

\bibitem[]{}Bouchy, F., Carrier, F., 2001, A\&A, 374, 5

\bibitem[]{}Carrier, F., Bouchy, F., Kienzle, F., Bedding, T.R., 
Kjeldsen, H., Butler, R.P., Baldry, I.K., O'Toole, S.J., Tinney, 
C.G., Marcy, G.W., 2001, A\&A, 378, 142

\bibitem[]{}Cunha, M.S., 2001, MNRAS, 325, 373

\bibitem[]{}Cunha, M.S., Gough, D., 2000, MNRAS, 319, 1020

\bibitem[]{}Dziembowski, W., Goode, P.R., 1996, ApJ, 458, 338

\bibitem[]{}Gough, D.O., 2001, Science, 291 (5512), 2325

\bibitem[]{}Heller, C.H., Kawaler, S.D., 1988, ApJL, 329, L43

\bibitem[]{}Kawaler, S.D., \& Bradley, P.A. 1994, ApJ, 427, 415

\bibitem[]{}Kleinman, S.J., Nather, R.E., \& Phillips, T., 1996, 
PASP, 108, 356

\bibitem[]{}Kurtz, D.W., 1982, MNRAS, 200, 807

\bibitem[]{}Kurtz, D.W., Matthews, J.M., Martinez, P., Seeman, J., 
Cropper, M., Clemens, J.C., Kreidl, T.J., Sterken, C., Schneider, H., 
Weiss, W.W., Kawaler, S.D., Kepler, S.O., van der Peet, A., Sullivan, 
D.J., and Wood, H.J., 1989, MNRAS, 240, 881

\bibitem[]{}Matthews, J.M., Kurtz, D.W., Martinez, P., 1999, ApJ, 
511, 422

\bibitem[]{}Tassoul M., 1980, ApJS, 43, 469

\bibitem[]{}Tassoul M., 1990, ApJ, 358, 313

\bibitem[]{}Winget, D.E., Nather, R.E., Clemens, J.C., Provencal, J., 
Kleinman, S.J., Bradley, P.A., Wood, M.A., Claver, C.F., Grauer, 
A.D., Hine, B.P., Hansen, C.J., Fontaine, G., Wickramasinghe, D.T., 
Achilleos, N., Marar, T.M.K., Seetha, S., Ashoka, B.N., O'Donoghue, 
D., Warner, B., Kurtz, D.W., Buckley, D.A., Vauclair, G., Chevreton, 
M., Dolez, N., Barstow, M.A., Solheim, J.E., Ulla, A. M., Kanaan, A., 
Kepler, S.O., Henry, G.A., and Kawaler, S.D., 1991, ApJ, 378, 326

\end{thebibliography}
\end{document}